# Impact of JD Bernal's Thoughts in the Science of Science upon China: Implications for Today's Quantitative Studies of Science

Yong Zhao[1], Jian Du[2] and Yishan Wu[3]

[1] Information Research Center, China Agricultural University, Beijing, China
[2] National Institute of Health Data Science, Peking University, Beijing, China
[3] Chinese Academy of Science and Technology for Development, Beijing, China

**Abstract**：John Desmond Bernal (1901-1970) was one of the most eminent scientists in molecular biology, and also regarded as the founding father of the Science of Science. His book *The Social Function of Science* laid the theoretical foundations for the discipline. In this article, we summarize four chief characteristics of his ideas in the Science of Science: the socio-historical perspective, theoretical models, qualitative and quantitative approaches, and studies of science planning and policy. China has constantly reformed its scientific and technological system based on research evidence of the Science of Science. Therefore, we analyze the impact of Bernal's Science-of-Science thoughts on the development of China's Science of Science, and discuss how they might be usefully taken still further in quantitative studies of science.

**Keywords**：J.D. Bernal, the Social Function of Science, the Science of Science, China's Science of Science, Quantitative Studies of Science

## 1 Introduction

The Science of Science, literally meaning *science connoisseurship* or s*cience studies* (Kokowski, 2014) can be defined as the use of scientific methodology to study science itself. For example, one can study how research is done and where improvements can be made. The Science of Science can also be considered as the self-consciousness of science.

Polish scholars were the first to recognize a need for a separate field concerned with Science-of-Science problems (Walentynowicz, 1982). In 1925, the Polish sociologist Znaniecki introduced the term *naukoznawstwo* (*Science of Science or Science Studies*) in his study *Przedmiot i zadania nauki o wiedzy* (Subject and Tasks of the Science of Knowledge). Ten years later M. Ossowska and S. Ossowski wrote another article, entitled *Nauka o nauce* (*The Science of Science*) in Polish, in which they defined the scope and formulated the program of Science-of-Science research. These authors identified three core disciplines that constitute the scientific study of science: epistemology and philosophy of science, psychology of scientific creativity, and the anthropology or sociology of science.

The emergence of the Science of Science, as a new scientific discipline, is generally associated with J.D. Bernal's book *The Social Function of Science* first published in 1939 (Goldsmith & Mackay, 1964). The book's theme was summarized in its sub-title: "what science does and what science could do". In his later essay

*Towards a Science of Science*, co-authored with Mackay A.L., Bernal took Derek John de Solla Price's definition as a general indication of the discipline: "the history, philosophy, sociology, psychology, economics, political science and operations research (etc.) of science, technology, and medicine (etc.)". He also identified some specific research subjects in the Science of Science: statistical attacks, detailed study of critical cases, systems research, experimental approaches and classification. Meanwhile he separated the discipline into pure and applied branches (Bernal & Mackay, 1966). The first, descriptive and analytic in nature, asks "how do science and the scientist work?" and the second, normative and synthetic in nature, asks "how can science be applied to the needs of human society?" Bernal considered that the Science of Science must be a proper science with special characteristics. There must be observation, speculation and experiment or operational research. Thus, unlike Polish researchers in the history, philosophy and sociology of science, who approached Science-of-Science issues usually with a humanistic methodology, Bernal brought measurement techniques from statistics to the analysis of science (Price, 1964). He also pointed out emphatically that science is both affecting and being affected by the social changes of its time. He thought that "this is a social and economic rather than a philosophical inquiry" (Bernal, 1939).

Bernal is regarded as the founding father of the Science of Science (Price, 1964). Since 1981, to honor Bernal and his pioneering work in the Science of Science, the Society for Social Studies of Science (4S) awards the Bernal Prize annually to scholars who have made a distinguished contribution to the field. In fact, the thoughts of many famous scholars can be traced back to the influence or inspiration by Bernal. The list of such scholars include: the economist Christopher Freeman; the historian and physicist Mintomo Yuasa; the biochemist, historian and sinologist Joseph Needham. Especially, Derek de Solla Price, who was the first recipient of the Bernal Award, noted in his acceptance speech that his work in Scientometrics[1] was partly inspired by Bernal. Furthermore, Eugene Garfield ascertained that Bernal also helped prepare his mind to the sensitivity that led him to the field of scientific information retrieval and its milestone-like by-product *the Science Citation Index* (SCI) (Garfield, 2007).

Some previous studies introduced the academic career of Bernal and his contributions to the world of science (Muddiman, 2003; Andrew, 2005; Sheehan, 2007). In this article, we summarize Bernal's main thoughts in the Science of Science and discuss how they might be usefully taken still further in quantitative studies of science. In the following sections, we consider the four chief characteristics of Bernal's thoughts in the Science of Science: the socio-historical perspective, theoretical models, qualitative and quantitative approaches, and studies of science planning and policy. Then, we will discuss the impact of Bernal's Science of Science thoughts on the development of the Science of Science in China.

---

[1] Scientometrics, defined as the quantitative studies of science (Elkana et al., 1978) or the "quantitative study of science, communication in science, and science policy" (Hess, 1997), has its roots in the 1950s and 1960s, and stems from the work of the historian of science Derek de Solla Price (e.g., Price, 1963; 1965) in parallel to the development of the Science Citation Indexes by Eugene Garfield (Garfield, 1955; 1963).

## 2 Characteristics of Bernal's Thoughts in the Science of Science

*2.1 Socio-Historical Perspective*

Bernal viewed science as a social activity, integrally tied to the whole spectrum of other social activities: economic, cultural, philosophical and political (Bernal, 1939; 1954). His monograph, *The Social Function of Science*, laid the theoretical foundations for the Science of Science (Goldsmith & Mackay, 1964). In this book, Bernal examined how science was organized in Britain and elsewhere; the efficiency of scientific research; science in education; science and war; and international science. These aspects were analyzed as quantitatively as possible with the statistics and evidence available. He then went on to examine the ways in which science could be improved: the training of the scientist; the reorganization of research; scientific communication; the funding of science; the strategy of scientific advance; science in the service of man; science and social transformation; and the social function of science. Until now, many of the research topics of this book are still focal to quantitative studies of science.

Later Bernal developed the historical analysis of science further in his multi-volume work *Science in History* (Bernal, 1954). He considered that "the progress of science has been anything but uniform in time and place" and "periods of rapid advance have alternated with longer period of stagnation and even of decay". He listed different aspects of science and took science as an institution, a method, a cumulative tradition of knowledge, a major factor in the maintenance and development of production, and one of the most powerful influences molding beliefs and attitudes to the universe and man.

In our opinion, Bernal's conceptualization of science can provide the intellectual framework for a more comprehensive description and understanding of the structure and dynamics of science. However, the appropriate approach to integrate qualitative theories with quantitative perspectives is a central question in quantitative studies of science (Leydesdorff, 1989). A true Science of Science will have to be theoretically grounded and practically useful. It should allow for repeatability, economy, mensuration, heuristics, and consilience (Wilson, 1998); this view is also emphasized by scientometrician Henry Small (1998). Thus, some theoretical questions about science-of-science issues, proposed by Bernal from a socio-historical perspective, should be addressed in the quantitative studies of science, such as the shifting of the world's science center, the efficiency of scientific research, the interaction of science with instruments or technology, and the relationship between science and culture.

*2.2 Theoretical Models*

Theoretical models are the most potent instruments for scientific thinking. Börner et al. (2012) defined a model of science as "a systematic description of an object or phenomenon that shares important characteristics with its real-world counterpart and supports its detailed investigation". Bernal's work *The Social Function of Science* has influenced many of the descriptive models reflecting about science in a scientific manner (Garfield, 2007). Bernal saw science as a growing pyramid or a branching tree, or as a set of raids by a guerrilla troop (Bernal, 1939); all of them being excellent metaphors. Means of pushing the usefulness of such models further is provided as a by-product of the citation-index method of handling scientific literature – an activity that owes much to the revolutionary suggestions made by Bernal for the solution of the information crisis (Price, 1964). The citations in publications form a network linking them all together in a complex fashion. Nowadays, the analysis of citation network has been an important tool for understanding science dynamics in quantitative studies of science. In fact, the structure of Bernal's book *The Social Function of Science* reads like a *what-to-be-modeled* list (Ginda et al., 2015). Examples are organization (*The existing organization of research in Britain*), scientific practices (*The efficiency of scientific research*), scientific careers (*The training of the scientist*) or globalization (*International science*).

Models of science facilitate a theoretical and/or empirical understanding of the structure and dynamics of science with predictive power, and they are validated according to the accuracy of their predictions. In 2012, Springer published a book titled *Models of Science Dynamics* (Scharnhorst et al., 2012), which provided a review of major models of science. This book discussed different science model types. In general, descriptive models of science can be found in the field of philosophy of science, history of science, sociology of science, and science and technology studies, while predictive models of science (computational and mathematical) are developed in scientometrics, bibliometrics, system dynamics, physics and mathematics. However, for quantitative studies of science, in designing action (especially research evaluation activities), a crucial task now involves the wise creation and use of clarifying descriptors, which are very often expressed as quantities (Espeland & Stevens, 2008; Espeland & Stevens, 1998). In their absence, decision making can wallow in a morass of incoherent data; but wise action must be aware also that in the presence of quantities it is all too easy to take the model for the reality, and to solve problems that are chosen for their model-able elegance rather than for their content (Ramirez et al., 2019). All evaluative scientometricians should keep Ramirez's words in mind.

*2.3 Qualitative and quantitative approaches*

The *Social Functions of Science* gathered a wide range of quantitative indicators as well as qualitative insights to argue for the systematic organization and the expanded utilization of scientific research in society (Bernal, 1939). In the late 1950s, Bernal went on to make an important intellectual contribution to the development of the Science of Science by arguing in a number of papers for the adoption of qualitative as well as quantitative methodological approaches (Muddiman, 2003). His paper published in the International Conference on Scientific Information (ICSI) held in Washington DC in 1958, *The Transmission of Information, A User's Analysis*, argued cogently for a "natural historical" approach to user studies which might "form a useful corrective to the predominantly quantitative and mechanical character of the conference" (Bernal, 1959). It is clear that Bernal favored methodologies that mix or combine both qualitative and quantitative approaches. Later on, the quantitative measurement was systematically used in the basically qualitative studies of science. Books such as *Towards a Metric of Science* (Elkana et al., 1978) presented a combination of qualitative and quantitative analysis of science. However, in recent decades, quantitative studies and qualitative studies seem to go different ways in science studies. Especially, with a growing dominance of computational approaches, a 'computational turn' has been advocated by some scholars in the Science of Science.

It is indeed very important to emphasize that the computational methods in the Science of Science should be accompanied by more thoroughgoing and focused qualitative investigations, characteristic of the early humanistic phases of the Science of Science research (Kawalec, 2019). In fact, to bridge the gap between qualitative and quantitative studies is also a grand challenge in the Science of Science. Mixed-methods designs[2], which seem to be the only adequate design for studying and solving complex problems and even 'wicked problems' (Churchman, 1967), would not be overemphasized for the future development of the Science of Science.

*2.4 Studies of science planning and policy*

Bernal is the main contributor to the theories about the planning of science (Bukharin et al., 2014). He considered that "perhaps a five- or ten- year scheme for the whole of science and shorter schemes for individual sciences would be workable and provision would have to be made for changes, as at any moment the important of new integrating discoveries might be such as to demand a complete recasting of pre-existing schemes" (Bernal, 1939). He emphasized the balance between fundamental research and applied research throughout any plan of scientific advance, and pointed out that the first stage in planning the general direction of scientific advance is a survey of existing knowledge and techniques in all departments of human life. Price (1964) suggested that the new understanding would advance with all the certainty and application of a science, and if it did the way was clear for rational and informed planning to replace opinion and expediency in the planning of science that Bernal and others foresaw as a necessity of modern civilization. Today, driven by big data sources, the Science of Science would produce many more exciting insights about the social processes that lead to scientific discovery. The planning of science will be strongly supported by data-driven predictions in the Science of Science (Clauset et al., 2017).

Anyone reading Bernal's *The Social Function of Science* today would immediately recognize it as a book about science policy. In fact, science policy analysis has been a regular part of the science studies with journals such as *Science and Public Policy* (founded in 1974) and *Research Policy* (founded in 1971). Meanwhile, there are many scholarly works building the link between the collection of data for S&T indicators and the science policy for promoting the growth of certain indictors, such as *Science and Technology Indicators: Their Use in Science Policy and Their Role in Science Studies*, edited by Van Raan (1988). Due to such duality, the

---

[2] Mixed methods research is an approach to inquiry that combines or associates both qualitative and quantitative forms. Mixed methods designs provide researchers, across research disciplines, with a rigorous approach to answering research questions (Aramo-Immonen, 2011).

Science of Science is neither 'hard' nor 'soft' science, neither is it 'natural' or 'social' science (Holbrook, 1992). While the qualitative analysis of science policy options may be regarded as a domain in social science, the analysis of S&T indicators falls very definitely into the area of the mathematical and statistical science. In short, we hold that Bernal's contributions in qualitative studies of science, such as his ideas concerning science planning and science policy, deserves full attention by science studies scholars in general, and by scholars in quantitative studies of science in particular.

## 3 Impacts of Bernal's Science of Science thoughts on the development of China's Science of Science

The development process of Science of Science in China was profoundly influenced by Bernal's thoughts about the Science of Science.

### 3.1 The institutionalization of China's Science of Science

To promote the Science of Science, Bernal encouraged the study of contemporary science as it happens by getting academic posts for the Science of Science (Bernal & Mackay, 1966). In China, we witnessed an institutionalization process for the Science of Science, including "getting academic posts for it". The most important milestone in the early formation of Science of Science as a discipline in China was the establishment of the Chinese Association for Science of Science and S&T Policy Research (CASSSP) in 1982. So far CASSSP has 4,464 registered members, including scholars, Ph.D students, research mangers, and government administrators for STI affairs. In recent years, there are more than one thousand participants in the annual academic conference held by CASSSP. Following Bernal's understanding of the discipline, CASSSP emphasizes both pure and applied research in the Science of Science since the pure research and applied research often feed into each other. At present, CASSSP consists of 20 Special Interest Groups (SIGs) in different research fields of the Science of Science, including SIGs on Theory of the Science of Science and Discipline Construction, S&T Policy, Technological Innovation, Scientometrics and Informetrics, S&T Evaluation, Entrepreneurship and Innovation, Technology Foresight, Policy Simulation, Human Resources for S&T, Science Communication and Popularization, Science and Economics, Public Management, Sociology of Science, S&T Project Management, Intellectual Property Policy, Commercialization of S&T Achievements, Regional Innovation, S&T Infrastructure, Science and Culture, and Civil-military Integration. Meanwhile, there are three Chinese academic journals in the Science of Science sponsored by CASSSP, including *Science Research Management* (founded in 1980), *Science of Science and Management of S&T* (founded in 1980), and *Studies in Science of Science* (founded in 1983). Furthermore, Science of Science courses have been offered at some Chinese universities since the 1980s. In the early 21st century, the Ministry of Education (MoE) of China issued a list of 100 must-read books for university students, including the translated Chinese version of *The Social Function of Science*. In the mid-1990s, the programs of

Master's degree and Ph.D degree in *Science of Science and Management of S&T* was approved by China's Academic Degree Commission of the State Council (ADCSC).

*3.2 Research in China's Science of Science*

Since the 1950s, many of Bernal's classic works have been translated and published in Chinese, which has a lasting promotion effect on the research in Science of Science. The list of such classic pieces include: *The Social Function of Science* (translated and published in 1950), *Towards a Science of Science* (translated and published in 1980), *Science in History* (translated and published in 1983), and *After Twenty-five Years* (translated and published in 1985). *Engels and Science* was translated in 2017 and distributed among the Science of Science scholars. It is noteworthy that the Chinese version of *The Social Function of Science* has been cited 1,938 times by Chinese authors alone in duxiu.com (16 February 2020), an index of Chinese books and articles, while Google Scholar indicates that the book has been cited 1,893 times by authors of the whole world.

Generally speaking, the Science of Science in China is organized into pure and applied branches as proposed by Bernal. The pure branch, aiming to facilitate scientific theories and methodologies for improved understanding of how science and the scientists work, mainly includes Sociology of Science and Scientometrics. Studies on the Sociology of Science and Scientometrics in China began in this same period, but then they went different ways in science studies. The theories and research traditions of famous scholars, such as John Desmond Bernal, Derek de Solla Price, Robert K. Merton, and Thomas S. Kuhn, are introduced and studied by Chinese scholars in the Sociology of Science, while scientometric research has been dominated by computational methods and information technology. In recent years, the methodological approach that linked scientometric methods with theoretical considerations is used for studying and solving complex problems in China, such as gender gap in science (Ma et al., 2018), transnational academic mobility (Li & Tang, 2019), and research integrity (Tang et al., 2019).

The applied branch, in turn, uses scientific theories and methodologies to develop strategies for using Science of Science to meet the needs of human society. Such explorations include Studies of Science Policy and Management, Legal Study of Science, and Study of Science Education. Since the 1990s, studies on technological innovation and STI policy have been emphasized in China's Science of Science community. In recent years, China's leaders have been emphasising that the strategy of innovation-driven development should be fully implemented, and that innovation has become the primary engine of social and economic development. The country has constantly reformed its scientific and technological system based on research evidence of the Science of Science.

Overall, in the last 40 years, Bernal's thoughts in Science of Science has been adsorbed and developed in China. Meanwhile, China's Science of Science research has been evolving from the relatively general study to its more applied fields (such as Study on Innovation Policy, Study of Science Ethics, and Study of Science Education), from the qualitative analysis to the mixed (qualitative and quantitative) analysis, and from the study on general social functions of science to the study of more specific economic functions and strategic functions of science.

*3.3 Prominent Chinese scholars in the Science of Science*

Many Chinese scholars were enthralled by the Science of Science as proposed by Bernal. Due to space limitation, here we only mention two representative Chinese scholars in the Science of Science. Mr. Hsue-shen Tsien (1911-2009), a prominent Chinese scientist, regarded as China's Father of Missiles, took the lead to initiate *Science of Science* in China (Liu, 2012) and published the first Chinese paper on the Science of Science (Tsien, 1979). Tsien considered that the Science of Science belongs to the social sciences, provides the theoretical foundation of the scientific system, and is situated at *Technological Sciences (Ji Shu Ke Xue)*[3] level in the social-science system. The Science of Science takes the entire scientific knowledge as research object, including three branches, Study of the S&T System, the Study of Science Capacity, and the Political Science of Science.

Mr. Hongzhou Zhao (1941-1997), one of the pioneers of the Science of Science as well as Scientometrics in China, explored the studies of Science Capacity. His monograph *Ke Xue Neng Li Xue Yin Lun* (Introduction to the Study of Science Capacity) was published in 1984. This study provided a systematic introduction to the elements of science capacities in a society and their interactions, and discussed the social function of groups of scientists, library and information systems, experimental technology systems, labor structure, and science education. Meanwhile, he further studied the shifting of the world's center of science as proposed by Bernal by using qualitative and quantitative analysis (Zhao & Jiang, 1985).

On the policy side, almost all the major designers of China's reform of science and technology system during the 1980s were the scholar-officials who were devotees to the Science of Science as proposed by Bernal. Their work did not only lay the theoretical foundations of China's Science of Science, but also promoted the formation and implementation of early S&T policies in China. For example, the establishment of the Youth Scientist Program by the National Natural Science Foundation of China (NSFC) was legitimated by Hongzhou Zhao's research evidence of scientists' social ages (Zhao & Jiang, 1986).

*3.4 S&T Planning in China Based on Research Evidence of the Science of Science*

Bernal's thought of science planning has been fully accepted and frequently emphasized in China. The Chinese government has made unremitting efforts to make and implement the national S&T plans since the late 1950s. We would like to just mention here that China has witnessed phenomenal progress in science, technology and innovation in recent two decades, as integral part of the so-called "Chinese Miracle". Robert Lawrence Kuhn, the Chairman of the Kuhn Foundation, summed up the six factors contributing to the "Chinese Miracle" (Kuhn, 2019).

---

[3] Hsue-shen Tsien (1979; 1983) considered that the whole system of science should be divided into natural science, social science, systems science, science of thinking, science of human bodies, and mathematical science. Each scientific area should also be divided into three levels, named basic sciences, technological sciences, and engineering technology.

One of the key factors is that the Chinese Government's policies and objectives are long-term, generally with long-term, medium and short-term goals, and policies and measures to achieve these goals are constantly adjusted and revised according to the situation (Kuhn, 2019). This long-term orientation is also reflected in science, technology and innovation (STI) plans in China. To better make STI plans at various levels (national, regional, urban, corporate, etc.), one needs sophisticated technology forecasting, foresight, prediction, and assessment, which are all attractive 'battle field' for ambitious scientometricians warriors.

In recent years, the continuation of technology foresight activities has nurtured a "foresight culture", which provides a stable, favorable, and "soft" environment for S&T planning. Since 2013, large-scale technology foresight activity, led by the Chinese national government, has been conducted by the Chinese Academy of Science and Technology for Development (CASTED), a think-tank under the Ministry of Science and Technology (MOST). This activity is usually implemented in three steps (technology evaluation, foresight survey, and key technology selection), and it adopts a combined qualitative and quantitative method using large-scale Delphi surveys and bibliometric analysis (Li et al., 2016). Further research as part of the exercise includes the key technology road-mapping, future scenarios making, and cross-impact and technology cluster analyses. The technology gap between China and the global advanced level has also been analysed in terms of both the overall S&T development status and some specific S&T domains, in order to make objective judgement about the true picture of science and technology in China. Such technology foresight exercises can make China's S&T planning more precise and accurate, because they helped decision-makers to understand future trends in S&T well and make optional policy responses promptly.

## 4 Conclusion and Discussion

Bernal was one of the founders of empirical research on science and the data-based study of organizational networks, material embodiments and operational mechanisms of contemporary science. The four chief characteristics of his thoughts in the Science of Science were analyzed in this article: socio-historical perspective, theoretical models, qualitative and quantitative approaches, and studies of science planning and policy. Historical experience has proved that adhering to Bernal's Science of Science thoughts is the driving forces for the development of the Science of Science in China, while the research in the Science of Science deeply impacted China's science, technology and innovation (STI) process, partly because many STI policy designers and STI activity mangers are influenced by the Science of Science in general, and by Bernal's ideas in particular.

Pitifully, it seems that many contemporary scholars in the Science of Science have failed to recognize the precious legacy of Bernal. Among 61 publications we retrieved on 16 September 2019 from Web of Science (WoS), whose titles contain 'science of science' and really means 'science of science' in the sense of 'science studies', only eight publications cited any work of Bernal, the founding father of the

Science of Science. Furthermore, just two scholars among the authors of the eight publications examined and evaluated the works of Bernal. One is Helena Sheehan, a historian of science at Dublin City University, who introduced Bernal's contributions to philosophy, politics and the science of science (Sheehan, 2007). The other is Eugene Garfield, who studied the impact of Bernal's book *The Social Function of Science* by using *HistCite* software (Garfield, 2009). No matter it is willful neglect or unintentional omission, such unfair or unwise behavior by the authors of the other 53 publications would be detrimental to the Science of Science itself.

Recently, two important literature reviews on the Science of Science were respectively published in *Physics Reports* and *Science*: *The Science of Science: From the Perspective of Complex Systems* (Zeng et al., 2017) and *Science of Science* (Fortunato et al., 2018). The common characteristic of these two different literature reviews is that the authors sensitively grasp the new characteristics and trends of contemporary science itself. Many of their ideas can be traced back to Bernal's thoughts, for instance, the ideas that science can be described as a complex, self-organizing, and evolving network of scholars, projects, papers, and ideas; that contemporary science is a dynamical system of undertakings driven by complex interactions among social structures, knowledge representations, and the natural world (Fortunato et al., 2018). Together with the development of science itself, the Science of Science has become an important research field. It seeks to understand, quantify and predict scientific research and the resulting outcomes (Zeng et al., 2017).

Today, the new scientometric analysis delineate important changes in science. It seems to us that scientometric analysis increasingly calls for a theoretical underpinning in order to explain and understand the mechanism behind the observed dynamics. Bernal's main thoughts, especially his theoretical explorations, might be usefully taken still further in quantitative studies of science.

**Acknowledgments**

This work was financially supported by the MOE (Ministry of Education in China) Project of Humanities and Social Sciences (18YJC870027). The authors gratefully thank Professor Loet Leydesdorff and Associate Professor Staša Milojević for the constructive comments and recommendations.